# Unconventional Charge-density-wave Order in a Dilute *d*-band Semiconductor

Huandong Chen[1,13], Batyr Ilyas[2,13], Boyang Zhao[1,13], Emre Ergecen[2,13], Josh Mutch[3], Gwan Yeong Jung[4], Qian Song[2], Connor A. Occhialini[2], Guodong Ren[5], Sara Shabani[6], Eric Seewald[6], Shanyuan Niu[1], Jiangbin Wu[7], Nan Wang[1], Mythili Surendran[1,8], Shantanu Singh[1], Jiang Luo[9], Sanae Ohtomo[3], Gemma Goh[10], Bryan C. Chakoumakos[11], Simon J. Teat[12], Brent Melot[1,10], Han Wang[7], Di Xiao[3], Abhay N. Pasupathy[6], Riccardo Comin[2], Rohan Mishra[4,5], Jiun-Haw Chu[3], Nuh Gedik[2], Jayakanth Ravichandran[1,7,8]*

[1]Mork Family Department of Chemical Engineering and Materials Science, University of Southern California, Los Angeles CA, USA

[2]Department of Physics, Massachusetts Institute of Technology, Cambridge MA, USA

[3]Department of Physics, University of Washington, Seattle WA, USA

[4]Department of Mechanical Engineering and Materials Science, Washington University in St. Louis, St. Louis MO, USA

[5]Institute of Materials Science & Engineering, Washington University in St. Louis, St. Louis MO, USA

[6]Department of Physics, Columbia University, New York NY, USA

[7]Ming Hsieh Department of Electrical and Computer Engineering, University of Southern California, Los Angeles CA, USA

[8]Core Center for Excellence in Nano Imaging, University of Southern California, Los Angeles, California, USA

[9]Department of Chemistry, Washington University in St. Louis, St. Louis MO, USA

[10]Department of Chemistry, University of Southern California, Los Angeles, California, USA

[11]Neutron Scattering Division, Oak Ridge National Laboratory, Oak Ridge TN, USA

[12]Advanced Light Source, Lawrence Berkeley National Laboratory, Berkeley CA, USA

[13]These authors contributed equally to this work        *e-mail: j.ravichandran@usc.edu

**Abstract**

Electron-lattice coupling effects in low dimensional materials give rise to charge density wave (CDW) order and phase transitions[1-4]. These phenomena are critical ingredients for superconductivity and predominantly occur in metallic model systems such as doped cuprates[5-7], transition metal dichalcogenides[8,9], and more recently, in Kagome lattice materials[10-13]. However, CDW in semiconducting systems, specifically at the limit of low carrier concentration region, is uncommon. Here, we combine electrical transport, synchrotron X-ray diffraction and optical spectroscopy to discover CDW order in a quasi-one-dimensional (1D), dilute *d*-band semiconductor, $BaTiS_3$, which suggests the existence of strong electron-phonon coupling. The CDW state further undergoes an unusual transition featuring a sharp increase in carrier mobility. Our work establishes $BaTiS_3$ as a unique platform to study the CDW physics in the dilute filling limit to explore novel electronic phases.

**Introduction and main text**

Charge density wave (CDW) is a periodic modulation of electron density, which is accompanied by periodic lattice distortions and is often observed in low-dimensional metals or semimetals[1,14,15]. The discovery of unconventional CDW order in strongly correlated superconductors such as doped cuprates[5,6,16,17] continues to challenge and expand our understanding of CDW physics and its connection to superconductivity. Recently, there is a growing interest in semiconducting materials that show CDW transitions, such as 1$T$-TiSe$_2$[18,19], Ta$_2$NiSe$_5$[20,21], and EuTe$_4$[22,23], whose origin presumably go beyond the classic Peierls' picture. They show phenomena such as electronic carrier-type switching[24], toroidal dipolar structures[21,25], and unconventional hysteretic transitions[22], and this has led to a vigorous debate over the mechanism of these CDW transitions[22,26]. Here, we report the discovery of unconventional CDW order and phase transitions in a non-degenerate semiconductor, BaTiS$_3$, which broadens the realm of CDW physics, especially in semiconducting materials. Upon cooling, we observe, first an increase and then a decrease of electrical resistivity as well as the activation energy across two phase transitions, which signals the emergence and suppression of the CDW state, respectively. Combining single-crystal X-ray diffraction (XRD) and optical probes, we provide direct experimental evidence of CDW order in the system and track down its evolution in both electronic and lattice degrees of freedom. Our experiments establish semiconducting BaTiS$_3$ as a new model system to study rich electronic phases and phase transitions associated with CDW in dilute filling.

**Transport signatures of CDW phase transitions in BaTiS$_3$**

Quasi-one-dimensional (quasi-1D) chalcogenide, BaTiS$_3$, is a small bandgap semiconductor ($E_g \sim 0.3$ eV) with a hexagonal crystal structure composed of 1D chains of TiS$_6$ octahedra, stacked between Ba chains (Fig. 1a). Recently, unusual physical properties such as giant optical anisotropy[27] and abnormal glass-like thermal transport properties[28] were reported in this material, which promotes questions over its electronic properties. We performed electrical transport measurements on bulk single crystals of $BaTiS_3$ along the chain axis (*c*-axis) (Fig. 1b). Here, we identify two different phase transitions from their hysteretic transport behavior. The striking feature of these transitions is their non-monotonic temperature dependence of resistivity. On cooling from room temperature, the resistivity increases, and undergoes a phase transition at 240 K that we refer to as Transition II. On further cooling, it continuous to increase till 150 K, after which it undergoes another transition featuring a sharp drop in resistivity, which we call Transition I. The thermal activation energy, $\Delta E$, across the transitions ($\Delta E \cong \delta(\ln(R))/\delta(1/k_BT)$) shows a clear increase from 75 meV to 153 meV at Transition II and then, a drop to 85 meV at Transition I during cooling cycle (Extended Data Fig. 6). The Transition II (240-260 K) hints at the emergence of a CDW state from a high temperature semiconducting phase; while at Transition I (150-190 K), the electrical resistivity reduces dramatically by about an order of magnitude upon cooling and the hysteretic transition window spans over 40 K. This transition towards a more conductive state and the low temperature suppression of CDW order are rather unique among known CDW systems and points to an unconventional origin.

To further understand these observations, we performed Hall measurements to study the evolution of carrier concentration and mobility across both the phase transitions (Fig. 1c). At room temperature, the electron concentration is ~$1.1 \times 10^{18}$ cm$^{-3}$, and it decreases monotonically as temperature is lowered ($n < 10^{15}$ cm$^{-3}$ at 100 K). We did not observe any change in the carrier type across the transitions. These observations confirm the non-degenerate nature of $BaTiS_3$, which possesses one of the lowest carrier densities among reported CDW compounds. Further, the Hall mobility is found to increase by about one order of magnitude below Transition I, which is directly associated with observed drop in electrical resistance, suggesting the formation of a higher mobility ground state from the CDW phase. All these transport observations are consistent with two phase transitions that lead to a sequence of electronic phases in BaTiS$_3$, starting from a high-temperature semiconducting phase that transitions to a CDW phase at intermediate temperatures, and finally to a unique higher mobility phase at low temperatures.

**Evidence for CDW evolution revealed by structural characterization**

The superlattice reflections in a diffraction pattern is one of the direct experimental evidences for CDW, which indicates periodic lattice distortions associated with the charge modulation[24,29,30]. We performed synchrotron XRD at three different temperatures to track the structural changes during phase transitions. Fig. 2a shows the corresponding precession map projected onto the *hk*2 reciprocal plane. A hexagonal array of reflection spots is observed at room temperature consistent with $P6_3cm$ space group ($a = b$ = 11.7 Å, $c$ = 5.83 Å). At 220 K, additional superlattice reflections appear at $h+1/2\ k+1/2\ 2$ in the precession image, which signals a change in the periodicity of the lattice ($a = b$ = 23.3 Å, $c$ = 5.84 Å) associated with Transition II ($P6_3cm$ to $P3c1$). The intensities of these $2 \times 2$ commensurate superstructure reflections are two orders of magnitude lower than that of the primary reflections and are signatures of the formation of CDW order below Transition II. The unit cell doubling takes place in the *a-b* plane, rather than along the chain *c*-axis, which is usually the case in other classic quasi-1D CDW systems such as NbSe$_3$[31] and BaVS$_3$[32]. On further lowering the temperature to 130 K, the superlattice peaks disappeared and a new set of reflections associated with a smaller $\frac{2}{\sqrt{3}} \times \frac{2}{\sqrt{3}}$ unit cell emerged ($a = b$ = 13.4 Å, c = 5.82 Å), which indicates the suppression of the CDW *via* Transition I ($P3c1$ to $P2_1$). The low temperature space group $P2_1$ is not a subgroup of $P3c1$, and we have observed the large thermal hysteresis from transport measurements, both of which lead us to the conclusion that the Transition I is a first-order transition.

To further gain insights into the evolution of electronic structures in BaTiS$_3$ across these transitions, we calculated the electronic band structure of the three phases, based on refined crystal structures from XRD, using density-functional theory (DFT) (Extended Data Fig. 9). As temperature is lowered, the bandgap of the system increases from 0.26 eV to 0.3 eV at Transition II and then drops to 0.15 eV across Transition I, which qualitatively agrees with the evolution of thermal activation barrier from Arrhenius analysis of transport data (Extended Data Fig. 6).

## Optical characterization of BaTiS$_3$

Raman spectroscopy is a sensitive probe to detect symmetry changes in the lattice[33], and to study phase transitions[34,35]. To complement the structural analysis with XRD, we carried out temperature-dependent Raman scattering measurements on BaTiS$_3$, as presented in Fig. 3a. A clear change in structure with the emergence of 20 new phonon modes near ~170 K (Transition I) was observed, while the effect from Transition II is subtle. The $\frac{2}{\sqrt{3}} \times \frac{2}{\sqrt{3}}$ lattice reconstruction observed from XRD coincides with the emergence of new zone-folded Raman modes (See Methods). Furthermore, upon heating, we observe the softening of the mode near 35 cm$^{-1}$ with an increase in its bandwidth (Fig. 3b). This mode survives until at least 165 K, above which it gets buried under Rayleigh peak. Since the amount of softening is substantial and its energy is relatively small (35 cm$^{-1}$), we identify it as the soft mode of Transition I.

Given the knowledge of the electronic and structural degrees of freedom in these transitions, we seek to examine the interaction between these two subsystems. Optical transient reflectivity (TR) measurements provide insights into the nature of electron-phonon coupling[36,37] and the changes in the energy gap. In this experimental scheme, an ultrashort near-infrared pulse (1.63 eV) generates electron-hole pairs in BaTiS$_3$, which consequently release this energy *via* relaxation mechanisms such as electron-electron and electron-phonon interactions in ultrafast timescales; the TR traces possess the fingerprints of these relaxation mechanisms. By analyzing the electronic decay rate in TR, and its temperature evolution (Fig. 3c), we observe a critical divergence of relaxation time near 240 K (Transition II). The critical slowing down usually indicates the opening of an energy gap and is associated with a flattening of the Landau free energy at a second-order transition point, which are consistent with a CDW phase transition[37,38].

**Discussion**

CDW transitions in other ternary hexagonal chalcogenides such as BaVS$_3$ and BaNbS$_3$ are well documented in the literature[39-41]. These compounds are metallic with a $d^1$ electronic configuration, whereas stoichiometric BaTiS$_3$ has a $d^0$ configuration with a nominally empty conduction band. The carrier concentrations measured by Hall measurements in BaTiS$_3$ is too low to support a Peierls-type CDW transition. Another surprising observation is the two-dimensionality of the lattice ordering vector that lies in the *a-b* plane (Fig. 2), whereas BaTiS$_3$ consists of 1D chains of face-sharing TiS$_6$ octahedra and shows giant uniaxial optical anisotropy[27] at room temperature. To understand the underlying electronic anisotropy, we performed in-plane conductivity anisotropy measurements on BaTiS$_3$ (Extended Data Fig. 7). The measured anisotropy in resistivity ($\rho_a/\rho_c$) was ~ 4 based on Montgomery analysis[42,43], which is relatively small compared to other model quasi-1D metallic CDW systems like NbSe$_3$ ($\rho_a/\rho_c$ ~ 15-20)[44] and (TaSe$_4$)$_2$I ($\rho_a/\rho_c$ > 200)[45]. This suggests an important role of interchain coupling in stabilizing the CDW phase.

One of the proposed mechanisms for archetypical $d^0$-type semiconducting systems such as 1$T$-TiSe$_2$[18] and Ta$_2$NiSe$_5$[20] is the excitonic insulator state. As we did not observe any carrier type switching in BaTiS$_3$ from Hall measurements, we can rule out the role of electron-hole coupling in stabilizing the observed CDW order. Detailed analysis of the single crystal XRD at room temperature reveals anomalous Ti-displacements at room temperature with an antipolar structure, similar to 1$T$-TiSe$_2$[25] and Ta$_2$NiSe$_5$[21]. Hence, it is likely that the CDW transition is lattice driven with the dimerization of this underlying antipolar structure. Further studies are necessary to probe the evolution of phonon dispersion across the transitions to confirm the role of lattice as the driving force.

Moreover, BaTiS$_3$ is a narrow bandgap semiconductor with non-degenerate, dilute concentration of electrons. Hence, the role of electron-electron interaction could be non-trivial unlike most metallic or semi-metallic CDW systems. One way to evaluate the electron correlation effects in a material is by calculating the dimensionless parameter $r_s = \sqrt[3]{\frac{1}{\frac{4}{3}\pi n}} / \left(\frac{4\pi\varepsilon\hbar^2}{m_b e^2}\right)$, defined by the ratio of Wigner-sphere radius (dominated by electron-electron interaction) to Bohr radius for a 3D electronic system[46]. Using the experimentally measured carrier concentration $n$ and calculated band effective mass $m_b$ and static dielectric constant $\varepsilon$ from DFT, we obtained a $r_s$ value of BTS close to 4.4 at 200 K ($n = 8.2 \times 10^{16}$ cm$^{-3}$), and it further reaches 18.5 at 90 K ($n = 1.7 \times 10^{14}$ cm$^{-3}$). As systems with $r_s$ values ~ 100 are considered to be in the strongly correlated region[46], one cannot rule out the role of electronic correlations in BaTiS$_3$ in effecting these transitions. Further, a cooperative mechanism with electron-lattice and electron-electron interactions could also lead to the phase transitions in BaTiS$_3$.

To fully understand the origin of these transitions, we need to probe the electronic structure and phonon dispersion across the two transitions. Further studies on the response of the phase transitions in BaTiS$_3$ to external fields such as pressure, strain, and doping, and whether electronic phases such as superconductivity can emerge in BaTiS$_3$ are also of interest. These approaches may also provide a pathway to push the system further into the strongly correlated regime and/or even achieve novel polaronic correlated electronic phases[47,48] by manipulating the competition between electron-lattice and electron-electron interactions. Our experimental findings of CDW order with novel phase transitions expand the realm of CDW physics to semiconducting materials with dilute filling and establish BaTiS$_3$ as a unique platform featuring a rich phase diagram to study electron correlation and electron-lattice interactions.

## Acknowledgements

This work was supported by an ARO MURI with award number W911NF-21-1-0327, an ARO grant with award number W911NF-19-1-0137 and NSF grants with award numbers DMR-2122070 and 2122071. The work at MIT in Gedik Group (B.Y., E.E. and N.G.) was supported by the US Department of Energy, BES DMSE and Gordon and Betty Moore Foundation's EPiQS Initiative Grant GBMF9459. Work at MIT in the Photon Scattering Group (Q.S., C.A.O. and R.C.) was supported by the National Science Foundation under Grant No. 1751739. J.M., S.O. and J.-H.C. acknowledge the support of the David and Lucile Packard Foundation and the State of Washington funded Clean Energy Institute. This research used resources of the Advanced Light Source, which is a DOE Office of Science User Facility under contract No. DE-AC02-05CH11231. We acknowledge the use of computational resources from the Extreme Science and Engineering Discovery Environment (XSEDE), which is supported by NSF grant number ACI-1548562. A portion of this research used resources at the Spallation Neutron Source, a DOE Office of Science User Facility operated by the Oak Ridge National Laboratory. We thank Alfred Zong and Edoardo Baldini for fruitful discussions.

## Author contributions

H.C. and J.R. conceived the idea and designed the experiments. N.G. and R.C. supervised the Raman and transient reflectivity studies. H.C. fabricated the devices and performed electrical transport measurements. J.M., S.O., and G.G. contributed to the transport measurements. B.Z., S.J.T., and B.C.C. performed single crystal X-ray diffraction measurements. B.I., Q.S., and C.A.O. carried out Raman measurements. B.I. and E.E. performed transient reflectivity measurements. G.Y.J., G.R., and J.L. contributed to the theoretical calculations under the supervision of R.M. S.S. and E.S. performed STS studies under the supervision of A.N.P. S.N. and B.Z. grew $BaTiS_3$ crystals. All authors discussed the results. H.C., B.I., and J.R. wrote the manuscript with input from all other authors.

## Additional information

## Competing financial interests

# References

1   Grüner, G. The dynamics of charge-density waves. *Reviews of Modern Physics* **60**, 1129 (1988).

2   Johannes, M. & Mazin, I. Fermi surface nesting and the origin of charge density waves in metals. *Physical Review B* **77**, 165135 (2008).

3   Zhu, X., Guo, J., Zhang, J. & Plummer, E. Misconceptions associated with the origin of charge density waves. *Advances in Physics: X* **2**, 622-640 (2017).

4   Rossnagel, K. On the origin of charge-density waves in select layered transition-metal dichalcogenides. *Journal of Physics: Condensed Matter* **23**, 213001 (2011).

5   Torchinsky, D. H., Mahmood, F., Bollinger, A. T., Božović, I. & Gedik, N. Fluctuating charge-density waves in a cuprate superconductor. *Nature Materials* **12**, 387-391 (2013).

6   Gerber, S. *et al.* Three-dimensional charge density wave order in $YBa_2Cu_3O_{6.67}$ at high magnetic fields. *Science* **350**, 949-952 (2015).

7   Ishihara, S. & Nagaosa, N. Interplay of electron-phonon interaction and electron correlation in high-temperature superconductivity. *Physical Review B* **69**, 144520 (2004).

8   Sipos, B. *et al.* From Mott state to superconductivity in 1$T$-$TaS_2$. *Nature Materials* **7**, 960-965 (2008).

9   Ma, L. *et al.* A metallic mosaic phase and the origin of Mott-insulating state in 1$T$-$TaS_2$. *Nature Communications* **7**, 1-8 (2016).

10  Neupert, T., Denner, M. M., Yin, J.-X., Thomale, R. & Hasan, M. Z. Charge order and superconductivity in kagome materials. *Nature Physics*, 1-7 (2021).

11  Chen, H. *et al.* Roton pair density wave in a strong-coupling kagome superconductor. *Nature* **599**, 222-228 (2021).

12   Zhao, H. *et al.* Cascade of correlated electron states in the kagome superconductor CsV$_3$Sb$_5$. *Nature* **599**, 216-221 (2021).

13   Teng, X. *et al.* Discovery of charge density wave in a correlated kagome lattice antiferromagnet. *arXiv preprint arXiv:2203.11467* (2022).

14   Peierls, R. & Peierls, R. E. *Quantum Theory of Solids*. (Oxford University Press, 1955).

15   Thorne, R. E. Charge-density-wave conductors. *Physics Today* **49**, 42-47 (1996).

16   Achkar, A. *et al.* Orbital symmetry of charge-density-wave order in La$_{1.875}$Ba$_{0.125}$CuO$_4$ and YBa$_2$Cu$_3$O$_{6.67}$. *Nature Materials* **15**, 616-620 (2016).

17   Frano, A. *et al.* Long-range charge-density-wave proximity effect at cuprate/manganate interfaces. *Nature Materials* **15**, 831-834 (2016).

18   Cercellier, H. *et al.* Evidence for an Excitonic Insulator Phase in 1$T$−TiSe$_2$. *Physical Review Letters* **99**, 146403 (2007).

19   Kidd, T., Miller, T., Chou, M. & Chiang, T.-C. Electron-hole coupling and the charge density wave transition in TiSe$_2$. *Physical Review Letters* **88**, 226402 (2002).

20   Lu, Y. *et al.* Zero-gap semiconductor to excitonic insulator transition in Ta$_2$NiSe$_5$. *Nature Communications* **8**, 1-7 (2017).

21   Nakano, A. *et al.* Antiferroelectric distortion with anomalous phonon softening in the excitonic insulator Ta$_2$NiSe$_5$. *Physical Review B* **98**, 045139 (2018).

22   Lv, B. *et al.* Unconventional hysteretic transition in a charge density wave. *Physical Review Letters* **128**, 036401 (2022).

23   Wu, D. *et al.* Layered semiconductor EuTe$_4$ with charge density wave order in square tellurium sheets. *Physical Review Materials* **3**, 024002 (2019).

24	Di Salvo, F. J., Moncton, D. & Waszczak, J. Electronic properties and superlattice formation in the semimetal TiSe$_2$. *Physical Review B* **14**, 4321 (1976).

25	Kitou, S. *et al.* Effect of Cu intercalation and pressure on excitonic interaction in 1$T$−TiSe$_2$. *Physical Review B* **99**, 104109 (2019).

26	Rossnagel, K., Kipp, L. & Skibowski, M. Charge-density-wave phase transition in 1$T$−TiSe$_2$: Excitonic insulator versus band-type Jahn-Teller mechanism. *Physical Review B* **65**, 235101 (2002).

27	Niu, S. *et al.* Giant optical anisotropy in a quasi-one-dimensional crystal. *Nature Photonics* **12**, 392-396 (2018).

28	Sun, B. *et al.* High frequency atomic tunneling yields ultralow and glass-like thermal conductivity in chalcogenide single crystals. *Nature Communications* **11**, 1-9 (2020).

29	Tsutsumi, K. *et al.* Direct electron-diffraction evidence of charge-density-wave formation in NbSe$_3$. *Physical Review Letters* **39**, 1675 (1977).

30	Moncton, D., Axe, J. & DiSalvo, F. Study of superlattice formation in 2$H$-NbSe$_2$ and 2$H$-TaSe$_2$ by neutron scattering. *Physical Review Letters* **34**, 734 (1975).

31	Hodeau, J. *et al.* Charge-density waves in NbSe$_3$ at 145K: Crystal structures, X-ray and electron diffraction studies. *Journal of Physics C: Solid State Physics* **11**, 4117 (1978).

32	Inami, T. *et al.* Symmetry breaking in the metal-insulator transition of BaVS$_3$. *Physical Review B* **66**, 073108 (2002).

33	Zhang, K. *et al.* Raman signatures of inversion symmetry breaking and structural phase transition in type-II Weyl semimetal MoTe$_2$. *Nature Communications* **7**, 1-6 (2016).

34	Xi, X. *et al.* Strongly enhanced charge-density-wave order in monolayer NbSe$_2$. *Nature Nanotechnology* **10**, 765-769 (2015).

35   Goli, P., Khan, J., Wickramaratne, D., Lake, R. K. & Balandin, A. A. Charge density waves in exfoliated films of van der Waals materials: evolution of Raman spectrum in TiSe$_2$. *Nano Letters* **12**, 5941-5945 (2012).

36   Gadermaier, C. *et al.* Electron-phonon coupling in high-temperature cuprate superconductors determined from electron relaxation rates. *Physical Review Letters* **105**, 257001 (2010).

37   Demsar, J., Biljaković, K. & Mihailovic, D. Single particle and collective excitations in the one-dimensional charge density wave solid K$_{0.3}$MoO$_3$ probed in real time by femtosecond spectroscopy. *Physical Review Letters* **83**, 800 (1999).

38   Lin, T. *et al.* Optical spectroscopy and ultrafast pump-probe study on Bi$_2$Rh$_3$Se$_2$: Evidence for charge density wave order formation. *Physical Review B* **101**, 205112 (2020).

39   Nakamura, M. *et al.* Metal-semiconductor transition and Luttinger-liquid behavior in quasi-one-dimensional BaVS$_3$ studied by photoemission spectroscopy. *Physical Review B* **49**, 16191 (1994).

40   Wada, T., Shimada, M. & Koizumi, M. Synthesis of Ba(V$_{1-x}$Ti$_x$)S$_3$ ($0 \leqq x \leqq 1.0$) compounds and their structural transitions. *Journal of Solid State Chemistry* **33**, 357-359 (1980).

41   Kim, S.-J. *et al.* Structure and physical properties of the barium niobium sulfides BaNbS$_3$ and BaNb$_{0.8}$S$_{3-\delta}$. *Journal of Solid State Chemistry* **115**, 427-434 (1995).

42   Montgomery, H. Method for measuring electrical resistivity of anisotropic materials. *Journal of applied physics* **42**, 2971-2975 (1971).

43   Borup, K. A., Fischer, K. F., Brown, D. R., Snyder, G. J. & Iversen, B. B. Measuring anisotropic resistivity of single crystals using the van der Pauw technique. *Physical Review B* **92**, 045210 (2015).

44   Ong, N. P. & Brill, J. Conductivity anisotropy and transverse magnetoresistance of NbSe$_3$. *Physical Review B* **18**, 5265 (1978).

45   Smontara, A., Tkalčec, I., Bilušić, A., Budimir, M. & Berger, H. Anisotropy of the thermal conductivity in (TaSe$_4$)$_2$I. *Physica B: Condensed Matter* **316**, 279-282 (2002).

46   Ceperley, D. M. & Alder, B. J. Ground state of the electron gas by a stochastic method. *Physical Review Letters* **45**, 566 (1980).

47   Rastelli, G. & Ciuchi, S. Wigner crystallization in a polarizable medium. *Physical Review B* **71**, 184303 (2005).

48   Fratini, S. & Quémerais, P. Polarization catastrophe in the polaronic Wigner crystal. *The European Physical Journal B-Condensed Matter and Complex Systems* **29**, 41-49 (2002).

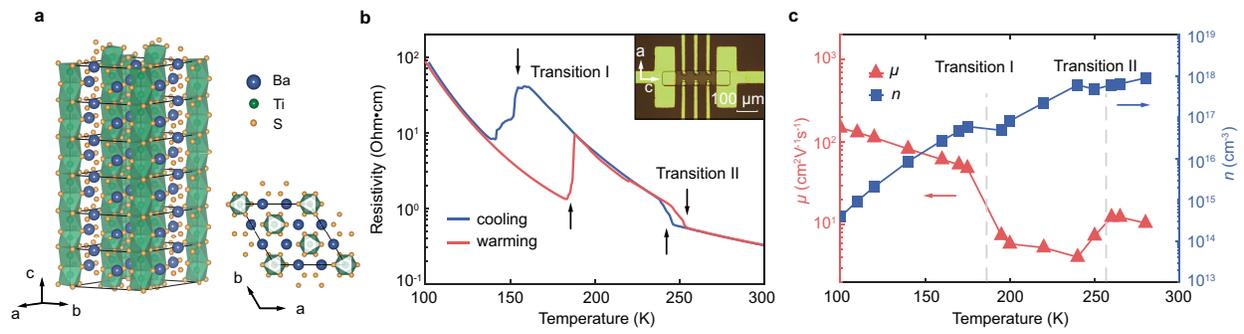

**Fig. 1 | Signature of phase transitions from electrical transport measurements**. **a**, Room temperature $P6_3cm$ structure of BaTiS$_3$, showing hexagonal symmetry. **b**, Illustration of representative temperature dependent electrical resistivity of BaTiS$_3$ crystal along c-axis. Abrupt and hysteric jumps in resistance are shown near 150-190 K (Transition I), and 240-260 K (Transition II). Inset shows optical microscopic image of a BaTiS$_3$ device used for Hall measurements. **c**, Temperature dependence of the mobility, $\mu$, and carrier concentration, $n$, of the dominant carrier, extracted from Hall measurements during a warming cycle.

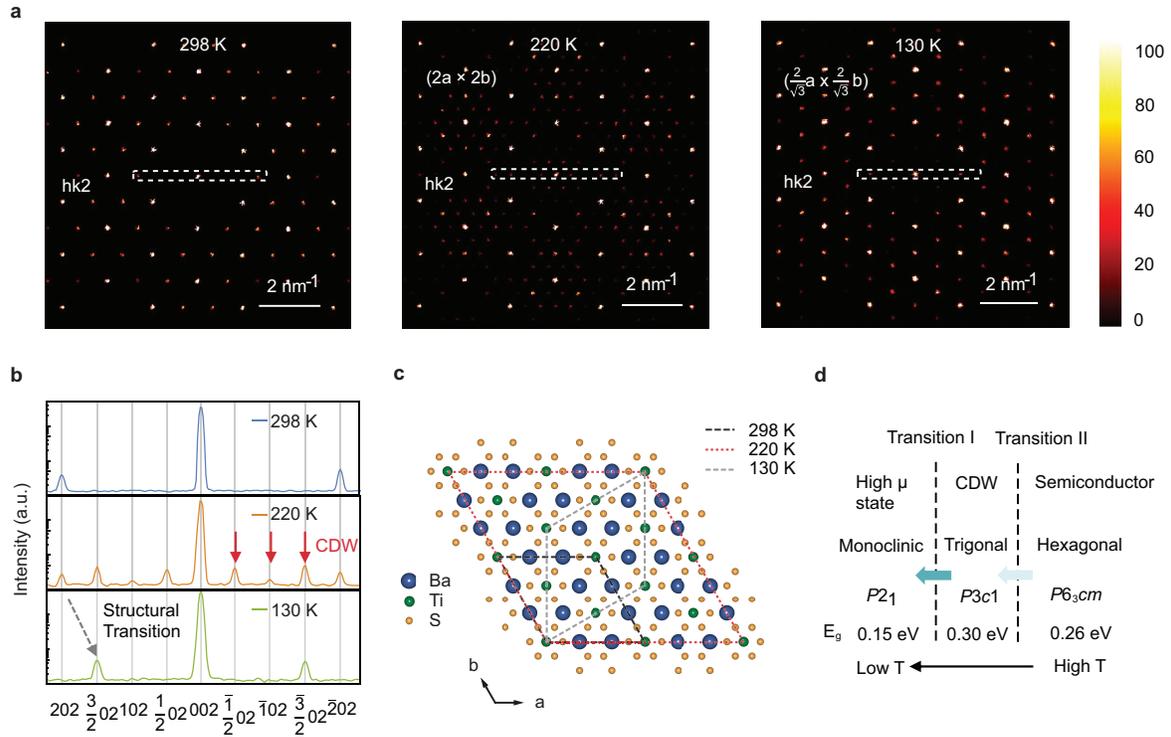

**Fig. 2 | Evidence for CDW order evolution via single crystal XRD**. **a**, Precession images from single-crystal X-ray diffraction measurements along hk2 projection at 298 K, 220 K and 130 K. **b**, X-ray intensity cut along the direction as indicated in Fig. 2**a**. **c**, Illustration of unit cell evolution of BaTiS$_3$ at different temperatures. **d**, Summary of electronic phases and phase transitions in BaTiS$_3$.

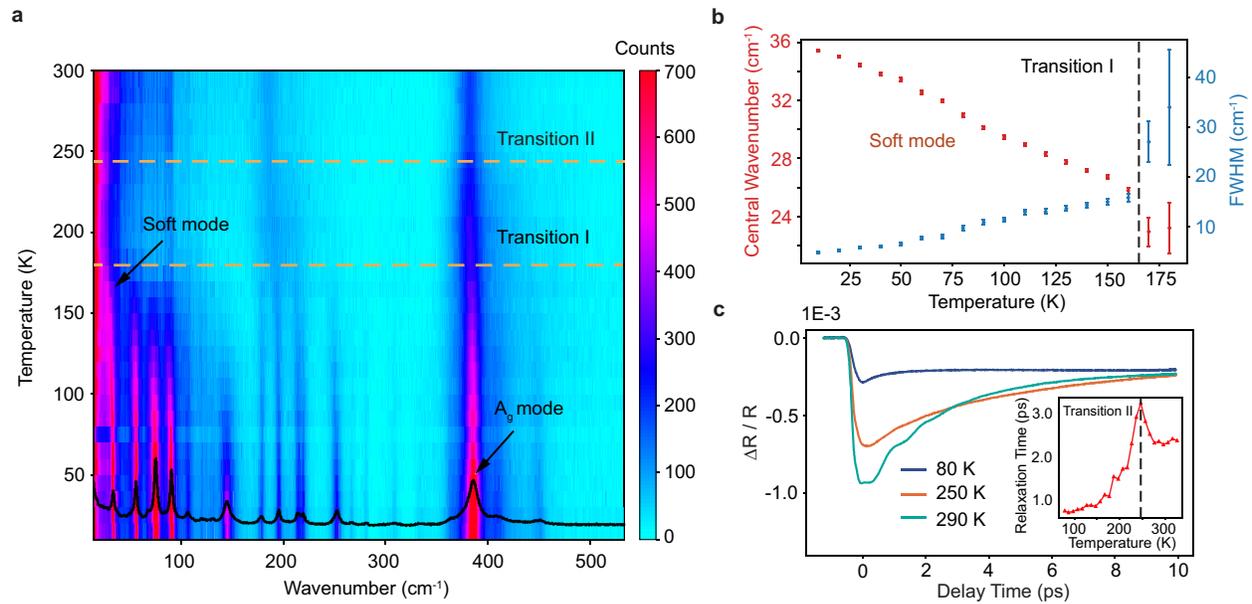

**Fig. 3 | Optical characterization of phase transitions. a**, Temperature dependent Raman spectra (parallel channel) of $BaTiS_3$. The spectrum shows a marked change upon cooling below ~170 K (Transition I), with more than 20 new Raman modes emerging. **b**, Mode hardening of the "Soft mode" near 35 cm$^{-1}$. The bandwidth (in blue) widens upon heating across the transition temperature. **c**, Transient reflectivity trace at three different temperatures (80 K, 250 K and 290 K). The relaxation times of excited carriers are extracted by fitting with a single exponential function plus a constant. The inset shows temperature dependence of relaxation times. Around ~240 K (Transition II), decay time diverges from both sides of this temperature.

## Methods:

**Crystal growth and room temperature characterization**

Single crystals of BaTiS$_3$ were grown by chemical vapor transport method as reported elsewhere[1]. Extended Data Fig.1 illustrates different morphologies of the obtained crystals with a- and c-axis in plane. Single-crystal diffraction performed at room temperature shows no substantial differences structurally between them (Extended Data Table 4). Needle-like crystals with well-defined c-axis are usually picked for transport measurements. For in-plane transport anisotropy measurements, as well as all temperature-dependent optical and single-crystal XRD characterization in this manuscript, we used BTS platelet samples.

Scanning tunneling microscopy (STM) / scanning tunneling spectroscopy (STS) measurements were performed on freshly cleaved surface from a needle sample at room temperature (Extended Data Fig. 2). Surface morphology scan clearly resolves the crystal surface with RMS roughness less than 0.5 nm, and dI / dV spectra indicates a semiconducting feature with ~ 0.3 eV bandgap, which is consistent with both the reported optical absorption spectra[1] and DFT calculations.

**Device fabrication**

As-grown BTS crystals (10-20 μm thick) are embedded in a polymeric media to planarize the top surface such that regular lithography and metallization methods can be readily applied, as illustrated in Extended Data Fig. 3a. Low stress polyimide with coefficient of thermal expansion (CTE) ~ $3 \times 10^{-6}$ K$^{-1}$ is chosen to minimize the effect of thermal contraction on transport behaviors at low temperatures. Ohmic contacts to BaTiS$_3$ crystals were fabricated using photolithography, followed by e-beam evaporation of Ti / Au (3 / 300 nm) and lift-off. An SF$_6$ / Ar RIE treatment step (10 / 40 sccm, 100 W, 100 mTorr, 1 min) was applied right before metal deposition to remove surface oxides and further reduce the contact resistance. Optical images of BTS devices are shown in Extended Data Fig. 3b and 3c.

**Electrical transport measurements and analysis**

Regular transport measurements were carried out in a JANIS 10 K closed-cycle cryostat from 100 K to 300 K. Standard low-frequency ($f$ = 17 Hz) AC lock-in techniques (Stanford Research SR830) were used to measure sample resistance in four-probe geometry, with an excitation current of about 100 nA. Qualitatively reproducible transport behavior showing two characteristic phase transitions was obtained from various devices, different cooling rate, as shown in Extended Data Fig. 4. Voltage-tuned transport measurements were performed with DC filed ranging from 0.9 V / mm to 15.1 V / mm, and the transition temperatures of Transition II were found to be reduced as DC bias increases (Extended Data Fig. 5).

In-plane conductivity anisotropy was assessed by Montgomery method[2,3] (Extended Data Fig. 7a and 7b). The analysis was carried out by using a van der Pauw geometry to calculate the conductivity anisotropy $\rho_a/\rho_c$ using $\left(\frac{l}{w}\right)^2 \cdot B_{\alpha,k}^2$, where $l$ and $w$ are sample dimensions and $B_{\alpha,k}$ is the conformal mapping function related to the ratio of van der Pauw resistance along different directions ($R_{aa}/R_{cc}$). The extracted anisotropy value is close to 4 at room temperature.

Hall measurements were performed in a PPMS (Quantum Design) equipped with a 14 T magnet. AC current was generated by a lock-in amplifier and passed through the device, $V_{xx}$, $V_{xy}$ as well as the current $I_{AC}$ were recorded simultaneously. Carrier concentration $n$ and mobility $\mu$ were extracted assuming a single carrier model. Extended Data Fig. 8 shows supplemental data from magneto-transport measurement. Unlike many other metallic CDW systems such as 2H-NbSe$_2$[4] and 1T-TiSe$_2$[5], no carrier type switching behavior was observed across either of the two phase transitions, indicated by the plot of Hall voltage V$_{xy}$ vs. B throughout the whole temperature range.

**Single crystal X-ray diffraction**

Single crystal diffraction at 130 K, 220 K and 298 K were carried out on beamline 12.2.1 at the Advanced Light Source (ALS), Lawrence Berkeley National Laboratory. Crystals were mounted on Mi-TeGen Kapton loops (Dual Thickness MicroMounts™) and placed in a nitrogen cold stream on the goniometer head of a Bruker D8 diffractometer, equipped with a PHOTONII CPAD detector operating in shutter-less mode. Diffraction data were collected using monochromatic synchrotron radiation with a wavelength of 0.72880 Å using silicon (111) monochromator. A combination of $f$ and $\omega$ scans with scan speeds of 1 s per 2 degrees for the $f$ scans, and 1 s per 0.15 degree for the ω scans at 2θ = 0 and -20°, respectively. The precession map was generated by Bruker APEX 3 with resolution 1.5 Å and thickness 0.1 with the refined unit cell. Note that the refined space group is different from previously reported $P6_3mc$[1]. It is mainly attributed to the improved brightness and resolution using synchrotron radiation that allows the observation of weak reflections. Detailed crystallography data and refinement results at different temperatures are listed in Extended Data Table 5 – Table 11.

**Raman spectroscopy and Raman mode analysis**

Polarization-resolved Raman spectroscopy was performed in a backscattering geometry using a confocal microscope spectrometer (Horiba Evolution). The incident beam (532 nm) was focused on a spot of size of 1-2 μm by using a 50 X objective lens. Laser power was set to 2.0 mW. The integration time of the spectrometer was 5 min. The beam was linearly polarized, and a half-wave plate was placed before the objective for polarization rotation. The analyzer was placed in front of the spectrometer entrance and was set parallel/perpendicular to the incident polarization when measuring XX or XY channel respectively. For polarization dependence scans, the half-wave plate was rotated from 0° to 180° with a 5° step.

Based on temperature dependent XRD data, we assume the following space and point groups at different temperatures. $BaTiS_3$ undergoes two structural phase transitions in the temperature range from 4 K up to room temperature. From ~ 250 K (Transition II) to 300 K, a hexagonal structure with space group: $P6_3cm$ and point group: $C_{6v}$ (6$mm$). From ~ 160 K (Transition I) to ~ 250 K, trigonal structure with space group: $P3c1$ and point group: $C_{3v}$ (3$m$). Below ~ 160 K (Transition I), a monoclinic structure with space group: $P2_1$ and point group: $C_2$ (2).

Normal modes of the lattice form irreducible representations of the point group, therefore providing information about their Raman activity and symmetries in real space. The optical representations of the point groups above are as follows:

$C_{6v}$ (6$mm$): $\Gamma_{Raman-active} = 8A_1 + 15E_1 + 15E_2$ (acoustic modes are not included)

$C_{3v}$ (3$m$): $\Gamma_{Raman-active} = 59A_1 + 119E$ (acoustic modes are not included)

$C_2$ (2): $\Gamma_{Raman-active} = 59A + 58B$ (acoustic modes are not included)

The peak intensity of a Raman scattering from a given mode $Q_i$ is given by:

$$I_i \propto \left| \hat{e}_s \frac{\partial \tilde{\alpha}_i}{\partial Q_i} \hat{e}_i \right|^2 \quad (1)$$

where $\tilde{\alpha}_i$ is a polarizability tensor element, and $\frac{\partial \tilde{\alpha}_i}{\partial Q_i}$ is a Raman tensor element for the mode $Q_i$. $\hat{e}_s$ and $\hat{e}_i$ are polarization states of scattered and incident lights, respectively.

a. **Analysis of Raman modes at room temperature (T = 300 K)**

We first consider room temperature hexagonal crystal structure with $P6_3cm$ space group and $C_{6v}$ (6mm) point group. The Raman tensors are given by:

$$A_1 = \begin{pmatrix} a & 0 & 0 \\ 0 & a & 0 \\ 0 & 0 & b \end{pmatrix}, \quad E_1 = \begin{pmatrix} 0 & 0 & c \\ 0 & 0 & 0 \\ c & 0 & 0 \end{pmatrix} \text{ and } \begin{pmatrix} 0 & 0 & 0 \\ 0 & 0 & c \\ 0 & c & 0 \end{pmatrix},$$

$$E_2 = \begin{pmatrix} d & 0 & 0 \\ 0 & -d & 0 \\ 0 & 0 & 0 \end{pmatrix} \text{ and } \begin{pmatrix} 0 & -d & 0 \\ -d & 0 & 0 \\ 0 & 0 & 0 \end{pmatrix} \tag{2}$$

We worked in a backscattering geometry on a (010) surface of BaTiS$_3$. The incident light is along -y direction and the reflected light is along +y direction. In the parallel-polarization channel, we set $\hat{e}_s = \hat{e}_i = (\cos\theta, 0, \sin\theta)$, where $\theta$ is an angle from x-axis. In the cross-polarization channel $\hat{e}_i = (\cos\theta, 0, \sin\theta)$ and $\hat{e}_s = (-\sin\theta, 0, \cos\theta)$. From these we obtain polarization angle dependence of Raman peak intensities:

$$I(A_1) \propto |\hat{e}_s A_1 \hat{e}_i|^2 = [a(\cos\theta)^2 + b(\sin\theta)^2]^2 \quad (Paralell) \tag{3}$$

$$I(A_1) \propto |\hat{e}_s A_1 \hat{e}_i|^2 = (a-b)^2 \frac{(\sin 2\theta)^2}{4} \quad (Cross) \tag{4}$$

From polarization dependence of 385 cm$^{-1}$ peak, we assign it to $A_1$ mode.

The second Raman tensor matrix of $E_1$ will be silent in both channels of (010) surface. The intensity of the first matrix is given by:

$$I(E_1) \propto |\hat{e}_s E_1 \hat{e}_i|^2 = c^2(\sin 2\theta)^2 \quad (Parallel) \tag{5}$$

$$I(E_1) \propto |\hat{e}_s E_1 \hat{e}_i|^2 = c^2((\cos\theta)^2 - (\sin\theta)^2)^2 \quad (Cross) \tag{6}$$

The sum of intensities in two channels of $E_1$ is constant, independent of polarization angle. Thus, we assign 187 cm$^{-1}$ mode to $E_1$ mode.

Since $E_2$ modes are degenerate we add the intensities of two tensors.

$$I(E_2) \propto |\hat{e}_s E_2 \hat{e}_i|^2 = d^2(\cos\theta)^4 \quad (Parallel) \tag{7}$$

$$I(E_2) \propto |\hat{e}_s E_2 \hat{e}_i|^2 = \frac{d^2(sin2\theta)^2}{4} \quad (Cross) \quad (8)$$

We assigned 90 cm$^{-1}$ and 183 cm$^{-1}$ modes to $E_2$ symmetry. The results of room temperature mode analysis are summarized in Extended Data Table 1.

### b. Analysis of modes at low temperature (T = 15 K)

Next, we consider low temperature monoclinic structure of BaTiS$_3$ at T = 15 K. The assumed group is $P2_1$ and the point group is C$_2$ (2). The Raman tensor for this point group is given by:

$$A = \begin{pmatrix} a & d & 0 \\ d & b & 0 \\ 0 & 0 & c \end{pmatrix}, \quad B = \begin{pmatrix} 0 & 0 & e \\ 0 & 0 & f \\ e & f & 0 \end{pmatrix} \quad (9)$$

We worked in the backscattering geometry; thus, the Raman peak intensities are given by:

$$I(A) \propto |\hat{e}_s A \hat{e}_i|^2 = [a(\cos\theta)^2 + c(\sin\theta)^2]^2 \quad (Parallel) \quad (10)$$

$$I(A) \propto |\hat{e}_s A \hat{e}_i|^2 = (a-c)^2 \frac{(\sin 2\theta)^2}{4} \quad (Cross) \quad (11)$$

The modes that assigned to $A$ symmetry, are labeled in the table below (Extended Data Table 2).

$$I(B) \propto |\hat{e}_s B \hat{e}_i|^2 = e^2(\sin 2\theta)^2 \quad (Parallel) \quad (12)$$

$$I(B) \propto |\hat{e}_s B \hat{e}_i|^2 = e^2((\cos\theta)^2 - (\sin\theta)^2)^2 \quad (Cross) \quad (13)$$

The sum of Raman peak intensities in the parallel and cross channels of B mode, should be polarization independent. So, we sum the intensities of the two channels and assign polarization independent modes to B modes. Another way of distinguishing the A and B modes is as follows: 1) A mode in parallel channel must be two-fold, while the B mode must be four-fold. 2) in cross channel, both A and B must be four-fold, but with opposite phases. The results of these analysis are provided in the Extended Data Table 2. We observe 13 A modes and 11 B modes.

In Extended Data Fig. 10b and 10e, we plot polarization dependent Raman spectra in cross (XY) and in parallel (XX) channels respectively, at 15 K. Symmetry assignments of the modes are done by fitting each mode with expected polarization dependencies, from the Raman tensors of the assumed space and point groups. The soft mode (35 cm$^{-1}$) polarization angle dependence closely follows 385 cm$^{-1}$ with A symmetry (Extended Data Fig. 10c). Extended Data Fig. 10a shows temperature dependent Raman spectra in cross (XY) channel.

**Transient reflectivity measurement**

Experiments were performed with a Ti:sapphire oscillator lasing at the center wavelength of 760 nm (1.63 eV) and pulse duration of 25 fs was used. The repetition rate of the laser is 1.6 MHz and no cumulative heating effect was observed. The output was split into pump (higher intensity) and probe (low intensity) arms, with a translation stage placed on the probe path to induce time-delay between the two arms. To increase the signal to noise ratio of our setup, we modulate the pump intensity at 100 kHz. For faster data acquisition and averaging, the pump-probe delay is rapidly scanned at a rate of 5 Hz with an oscillating mirror. The reflected probe beam was sent directly to a photodiode. The signal from the photodiode is sent to a lock-in amplifier locked to the pump modulation frequency.

Tempearture dependent transient reflectivity measurements were performed from 80 K to 330 K. Signal traces at each tempreature consist of incoherent electronic decay signal and coherent phonon oscillations (Fig. 3d). We fit the decay background with a single exponential function:

$$S(t) = Ae^{-\frac{t}{\tau}} + B \tag{14}$$

where $\tau$ is a decay rate. The temperature dependence of the decay rate exhibits a cricital slowing down behavior, thus indicating an electronic phase tranition in BaTiS$_3$ near 240 K.

**DFT calculations**

The band structure of BaTiS$_3$ in the three different phases were computed using DFT. The initial structures were taken from the refined crystal structures from XRD at different temperatures, which are assigned to have a space group of $P6_3cm$ (298 K), $P3c1$ (220 K), and $P2_1$ (130 K), respectively. The structures were fully optimized by DFT. These calculations were done using the Vienna Ab initio Simulation Package (VASP)[6] with projector augmented wave (PAW) potentials[7]. The exchange-correlation energy was treated with the generalized gradient approximation (GGA) using the Perdew-Burke-Ernzerhof (PEB) functional[8]. A cutoff energy of 650 eV was used for the expansion of the plane waves. The convergence criteria were set to $10^{-8}$ eV for total energy and $10^{-4}$ eV/Å for atomic forces, respectively. The Brillouin zone was sampled using a $\Gamma$-centered Monkhorst-Pack $k$-points mesh[9], with maximum spacing of 0.05 Å$^{-1}$ for structural relaxations and 0.02 Å$^{-1}$ for static calculations, respectively. For the calculation of the $r_s$ parameter, density functional perturbation theory calculations[10] were performed to obtain the static dielectric constant ($\varepsilon$) with the same energy cutoff and $k$-point sampling as the DFT calculations. The input parameters for $r_s$, such as the band effective mass ($m_b^*$), experimental carrier concentration ($n$), and static dielectric constant ($\varepsilon$), are provided in the Extended Data Table 3.

1   Niu, S. *et al.* Giant optical anisotropy in a quasi-one-dimensional crystal. *Nature Photonics* **12**, 392-396 (2018).

2   Montgomery, H. Method for measuring electrical resistivity of anisotropic materials. *Journal of Applied Physics* **42**, 2971-2975 (1971).

3   Borup, K. A., Fischer, K. F., Brown, D. R., Snyder, G. J. & Iversen, B. B. Measuring anisotropic resistivity of single crystals using the van der Pauw technique. *Physical Review B* **92**, 045210 (2015).

4   Lee, H., McKinzie, H., Tannhauser, D. & Wold, A. The low‐temperature transport properties of NbSe$_2$. *Journal of Applied Physics* **40**, 602-604 (1969).

5   Di Salvo, F. J., Moncton, D. & Waszczak, J. Electronic properties and superlattice formation in the semimetal TiSe$_2$. *Physical Review B* **14**, 4321 (1976).

6   Kresse, G. & Furthmüller, J. Efficiency of ab-initio total energy calculations for metals and semiconductors using a plane-wave basis set. *Computational Materials Science* **6**, 15-50 (1996).

7   Blöchl, P. E. Projector augmented-wave method. *Physical Review B* **50**, 17953 (1994).

8   Perdew, J. P., Burke, K. & Ernzerhof, M. Generalized gradient approximation made simple. *Physical Review Letters* **77**, 3865 (1996).

9   Monkhorst, H. J. & Pack, J. D. Special points for Brillouin-zone integrations. *Physical Review B* **13**, 5188 (1976).

10  Baroni, S., De Gironcoli, S., Dal Corso, A. & Giannozzi, P. Phonons and related crystal properties from density-functional perturbation theory. *Reviews of Modern Physics* **73**, 515 (2001).

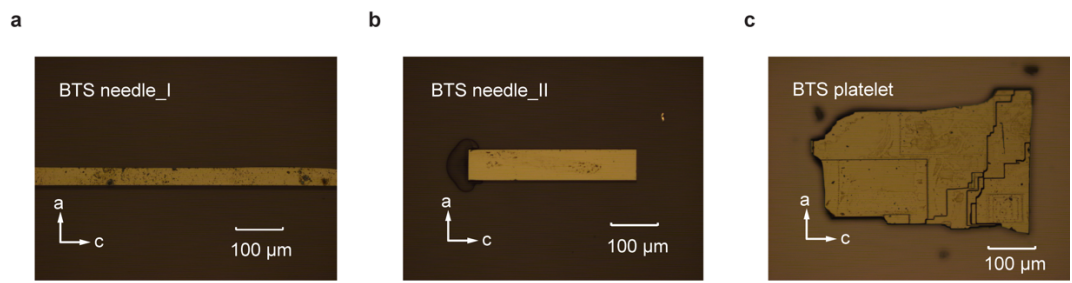

**Extended Data Fig. 1 | Different morphologies of CVT-grown BaTiS$_3$ crystals.** Optical images of **a**, thin BTS needle, **b**, thick BTS needle and **c**, BTS platelet. For 'needle'-like samples, c-axis follows the direction of long edges and is easy to determine. Edges of terraces on BTS plates are useful in assigning the crystal orientations, which can be further confirmed by polarization-dependent Raman spectroscopy.

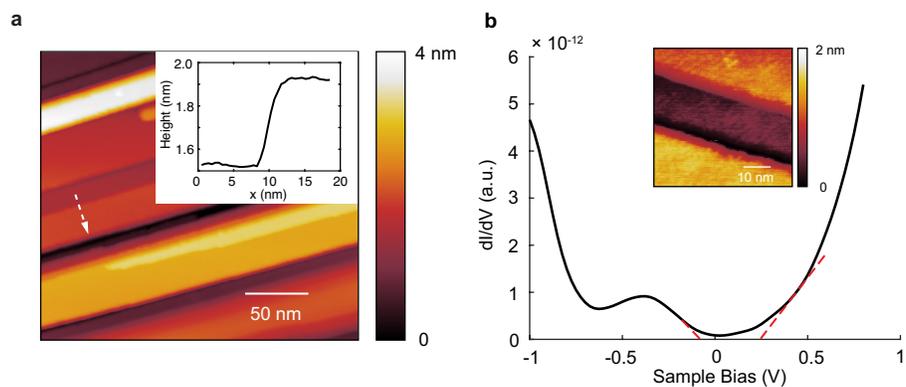

**Extended Data Fig. 2 | STM / STS study of BaTiS$_3$ at room temperature. a**, Surface morphology of freshly cleaved BaTiS$_3$ crystal. **b**, Typical dI/dV spectra on a freshly cleaved BaTiS$_3$ crystal surface taken at room temperature. A small bandgap of ~ 0.3 eV was extracted from the spectra, consistent with the value measured from optical absorption spectra.

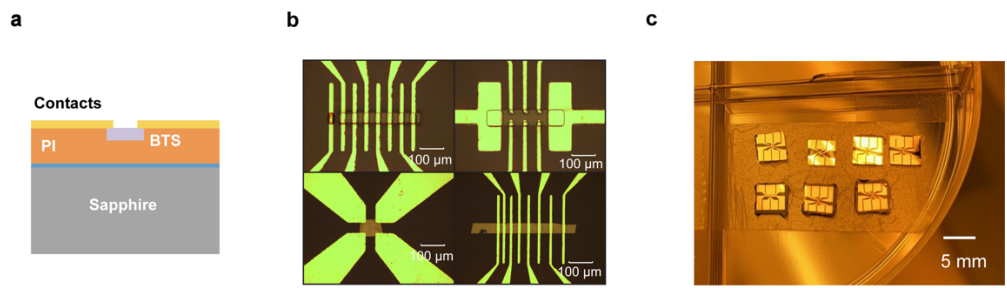

**Extended Data Fig. 3 | BaTiS$_3$ bulk device fabrication. a**, A cross-sectional illustration of a BaTiS$_3$ device for transport studies. Low-stressed polyimide is used for crystal embedding to minimize the extrinsic strain effect from thermal contraction. **b**, Optical microscopic images of BaTiS$_3$ devices with various electrode designs. **c**, Optical images of BaTiS$_3$ devices used for transport measurements.

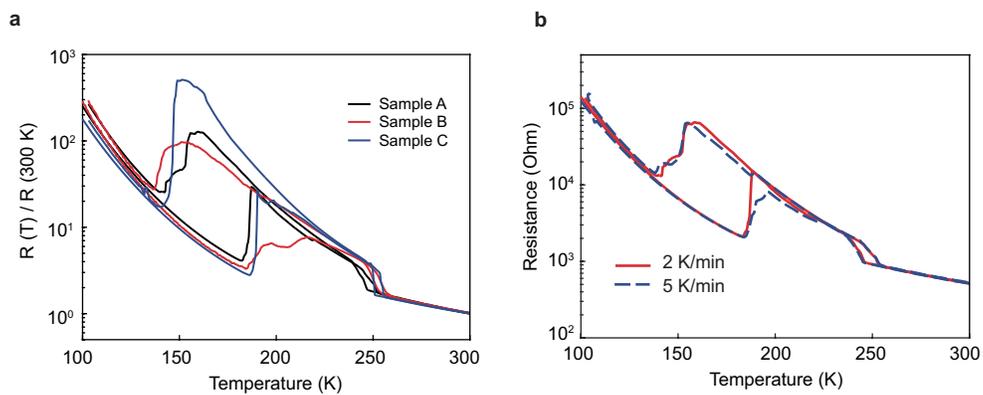

**Extended Data Fig. 4 | Reproducible transport measurements. a**, Temperature-dependent electrical resistance (normalized by R (300 K)) measured from three different BaTiS$_3$ devices. The transport behavior is qualitatively consistent with each other. **b**, Cooling rate-dependent transport on BaTiS$_3$. Neither of the transitions are suppressed or altered using cooling rate up to 5 K/min.

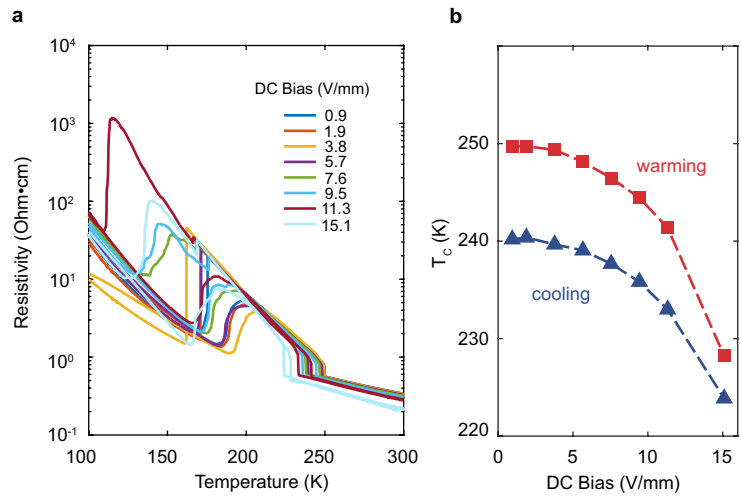

**Extended Data Fig. 5 | Voltage-tuned transport behavior. a**, Temperature-dependent electrical resistivity of BaTiS$_3$ crystal along c-axis measured at different DC voltage bias. **b**, Extracted transition temperature T$_C$ of Transition II (~ 250 K) as a function of DC bias for both cooling and warming cycles.

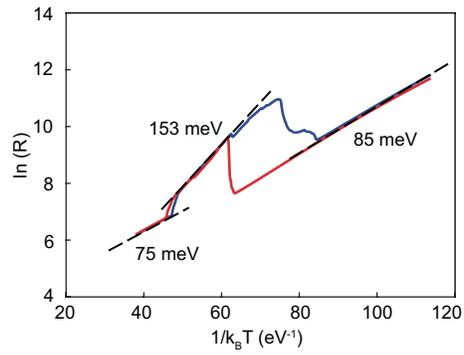

**Extended Data Fig. 6 | Thermal activation energy analysis of BaTiS$_3$ transport data presented in Fig. 1b.** The activation energy is obtained from $\Delta E = \frac{\delta(\ln(R))}{\delta(1/k_B T)}$ in the linear region. The extracted activation energy before and after Transition II, as well as after Transition I is 75 meV, 153 meV, and 85 meV, respectively.

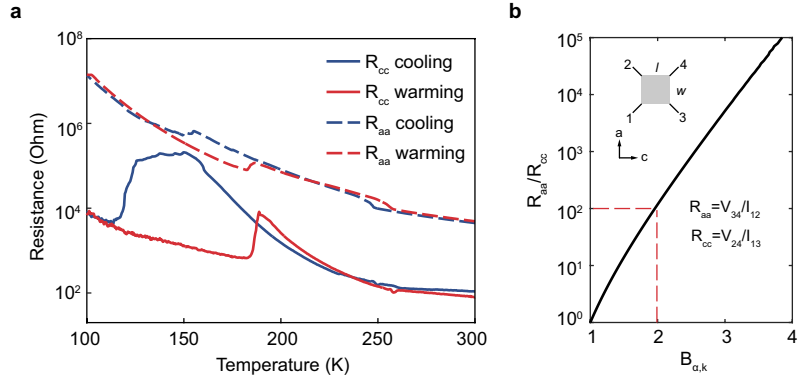

**Extended Data Fig. 7 | In-plane electrical conductivity anisotropy analysis using Montgomery method. a**, Measured van der Pauw resistance $R_{aa}$ and $R_{cc}$ as a function of temperature. The transport data was collected from a BaTiS$_3$ platelet device with van der Pauw geometry ($l = w = 20$ μm), as shown in the inset of Fig. S7 **b**. **b**, The conformal mapping function that uniquely maps the ratio $R_{aa}/R_{cc}$ to $B_{\alpha,k}$, which is later used to calculate the anisotropy $\rho_a/\rho_c$.

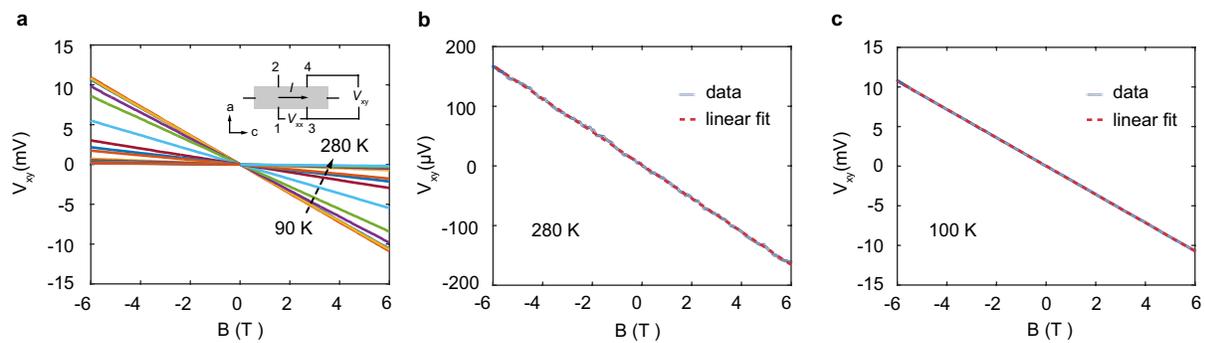

**Extended Data Fig. 8 | Hall measurements. a**, Plot of Hall voltage $V_{xy}$ with magnetic field from -6 T to 6 T at different temperatures. Hall data was collected from the Sample C, with its temperature-dependent longitudinal resistance illustrated in Extended Data Fig. 4a. **b-c**, The measured data (blue) were fitted linearly to extract the carrier concentration $n$ and mobility $\mu$.

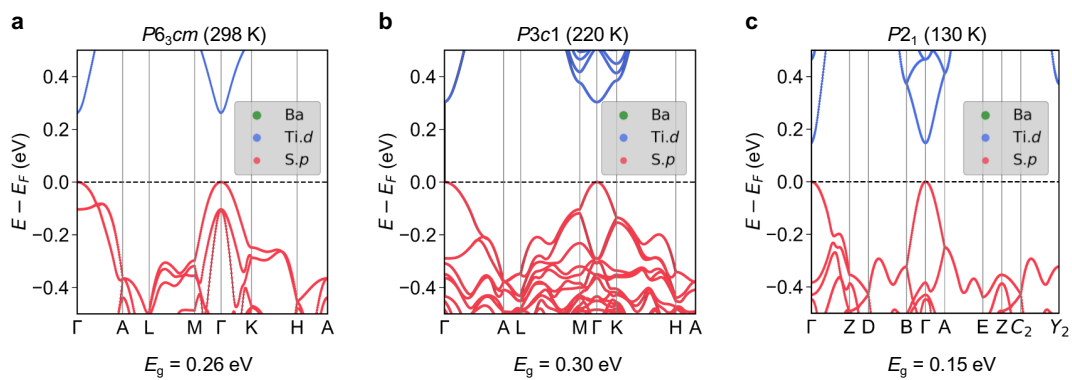

**Extended Data Fig. 9 | DFT-calculated electronic band structures of different phases of BaTiS$_3$. a**, $P6_3cm$, **b**, $P3c1$, and **c**, $P2_1$. The contribution of Ti $d$-states and S $p$-states to the band structure are highlighted with blue and red colors, respectively. The band gap ($E_g$) calculated at the PBE level are labeled below.

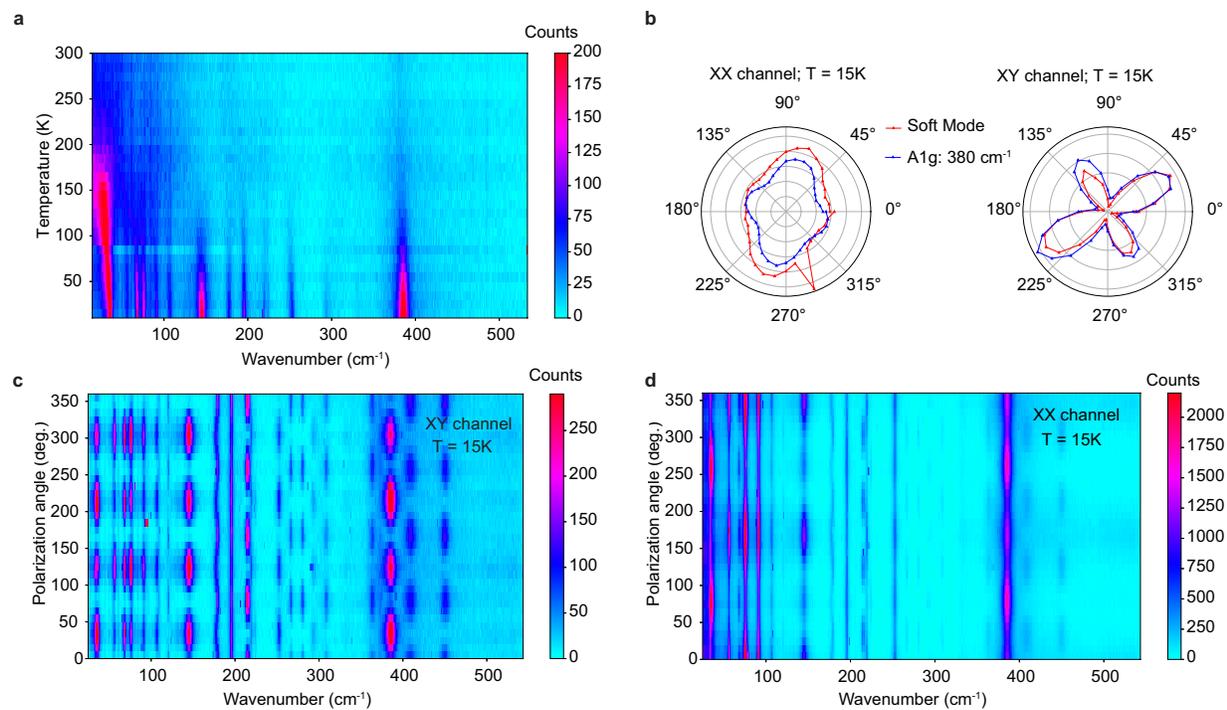

**Extended Data Fig. 10 | a**, Temperature dependent Raman spectra in cross channel. Like parallel channel, new modes emerge below Transition I. **b**, Polar plot of polarization dependence of soft mode and the $A_{1g}$ mode. The symmetry of soft mode in both channels coincides with the symmetry of $A_{1g}$ mode. **c**, Polarization dependence of cross channel Raman spectrum, taken at 15 K. **d**, Polarization dependence of parallel channel Raman spectrum, taken at 15 K.

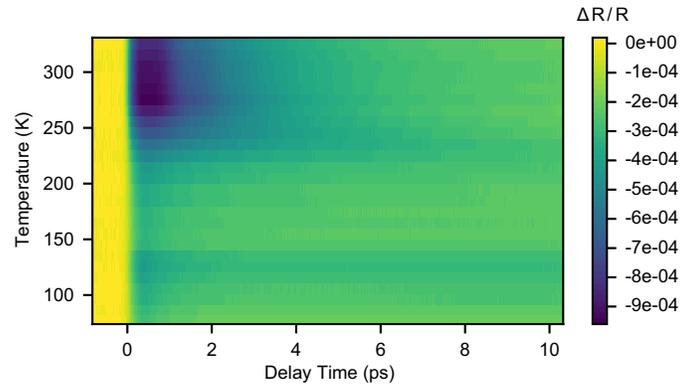

**Extended Data Fig. 11 | Contour plot of transient reflectivity as a function of temperature and delay time.**
The transient reflectivity traces at different temperatures shown in Fig. 3c are cut and taken from this plot.

**Extended Data Table 1. List of room temperature Raman modes (T = 300 K).**

| No. | Frequency (cm$^{-1}$) | Mode symmetry |
|---|---|---|
| 1 | 90 | $E_2$ |
| 2 | 183 | $E_2$ |
| 3 | 185 | $E_1$ |
| 4 | 385 | $A_1$ |

**Extended Data Table 2. List of low temperature Raman modes (T = 15 K).**

| No. | Frequency (cm$^{-1}$) | Mode symmetry |
|---|---|---|
| 1 | 35 | A |
| 2 | 56.5 | A |
| 3 | 68.5 | B |
| 4 | 76 | A |
| 5 | 91 | A |
| 6 | 106.5 | A |
| 7 | 120.5 | B |
| 8 | 145 | A |
| 9 | 177 | A |
| 10 | 179 | B |
| 11 | 196 | A |
| 12 | 214 | B |
| 13 | 217 | B |
| 14 | 219 | A |
| 15 | 224.5 | A |
| 16 | 253 | A |
| 17 | 266.5 | B |
| 18 | 280.5 | B |
| 19 | 293 | A |
| 20 | 309 | B |
| 21 | 363 | B |
| 22 | 385 | A |
| 23 | 410 | B |
| 24 | 450 | B |

**Extended Data Table 3. Input parameters used to compute the $r_s$ value for BaTiS$_3$.**

| Space group | $n$ (cm$^{-3}$) | $m_b$ | $\varepsilon_s$ | $r_s$ |
|---|---|---|---|---|
| $P2_1$ | $1.69 \times 10^{14}$ (90 K) | 0.522 | 59.64 | 18.56 |
| $P3c1$ | $8.18 \times 10^{16}$ (200 K) | 0.682 | 41.58 | 4.43 |
| $P6_3cm$ | $8.99 \times 10^{17}$ (280 K) | 0.539 | 35.16 | 1.86 |

**Extended Data Table 4. Data collection and refinement statistics for various BaTiS$_3$ morphologies at 300 K**

| | BaTiS$_3$ Platelet | BaTiS$_3$ Thin Needle | BaTiS$_3$ Thick Needle |
|---|---|---|---|
| **Space group** | $P6_3cm$ | $P6_3cm$ | $P6_3cm$ |
| **Cell dimensions** | | | |
| $a, b, c$ (Å) | 11.671(2), 11.671(2), 5.833(2) | 11.686(2), 11.686(2), 5.840(2) | 11.670(2), 11.670(2), 5.846(2) |
| $\alpha, \beta, \gamma$ (°) | 90.00(3), 90.00(3), 120.00(3) | 90.00(3), 90.00(3), 120.00(3) | 90.00(3), 90.00(3), 120.00(3) |
| Volume (Å$^3$) | 688.1(3) | 690.7(3) | 689.4(3) |
| Density (g/cm$^3$) | 4.075 | 4.060 | 4.067 |
| **Refinement** | | | |
| $R_1, wR_2$ [all data] | 0.0308, 0.0769 | 0.0341, 0.0762 | 0.0366, 0.0606 |
| $R_1, wR_2$ [$I > 4\sigma I$] | 0.0272, 0.0734 | 0.0326, 0.0747 | 0.0236, 0.0559 |
| GoF | 1.126 | 1.143 | 1.045 |
| Merohedral Twinning | -1 0 0  0 -1 0  0 0 -1 2 | -1 0 0  0 -1 0  0 0 -1 2 | -1 0 0  0 -1 0  0 0 -1 2 |
| BASF | 0.53047 | 0.39520 | 0.39818 |

**Extended Data Table 5. Data collection and refinement statistics of BaTiS$_3$ platelet at different temperatures**

|  | BaTiS$_3$ 300K | | BaTiS$_3$ 220K | | BaTiS$_3$ 130K | |
|---|---|---|---|---|---|---|
| **Space group** | $P6_3cm$ | | $P3c1$ | | $P2_1$ | |
| **Cell dimensions** | | | | | | |
| $a, b, c$ (Å) | 11.671(2), 11.671(2), 5.833(2) | | 23.285(5), 23.285(5), 5.836(2) | | 13.431(3), 5.817(2), 13.431(3) | |
| $\alpha, \beta, \gamma$ (°) | 90.00(3), 90.00(3), 120.00(3) | | 90.00(3), 90.00(3), 120.00(3) | | 90.00(3), 120.00(3), 90.00(3) | |
| Volume (Å$^3$) | 688.1(3) | | 2740.0(15) | | 908.8(5) | |
| Density (g/cm$^3$) | 4.075 | | 4.093 | | 4.114 | |
| **Intensity Statistics** | | | | | | |
| Resolution (Å) | Inf. - 0.62 | 0.72 - 0.62 | Inf. - 0.62 | 0.72 - 0.62 | Inf - 0.60 | 0.70 - 0.60 |
| Completeness (%) | 99.9 | 100.0 | 100.0 | 100.0 | 99.9 | 100.0 |
| Redundancy | 31.30 | 18.39 | 34.89 | 19.01 | 28.28 | 17.41 |
| Mean $I/\sigma I$ | 34.18 | 18.13 | 22.42 | 10.27 | 32.42 | 17.86 |
| **Reflections** | | | | | | |
| $R_\sigma, R_{int}$ | 0.0170, 0.0498 | | 0.0315, 0.0691 | | 0.0479, 0.0499 | |
| $\theta_{min}, \theta_{full}, \theta_{max}$ (°) | 2.015, 25.242, 34.703 | | 1.035, 25.930, 35.40 | | 1.751, 25.242, 36.160 | |
| Friedel fraction full | 0.990 | | 0.999 | | 0.999 | |
| **Refinement** | | | | | | |
| Resolution (Å) | ~ (Inf. - 0.62) | | ~ (Inf. - 0.62) | | 50.0 - 0.60 | |
| No. reflections | 1046 | | 7983 | | 8676 | |
| No. $I > 2\sigma I$ | 916 | | 6621 | | 7161 | |
| No. parameters | 30 | | 185 | | 187 | |
| No. constraints | 1 | | 1 | | 1 | |
| $R_1, wR_2$ [all data] | 0.0308, 0.0769 | | 0.0420, 0.0692 | | 0.0448, 0.0814 | |
| $R_1, wR_2$ [$I > 4\sigma I$] | 0.0272, 0.0734 | | 0.0305, 0.0657 | | 0.0321, 0.0764 | |
| GoF | 1.126 | | 1.058 | | 0.969 | |
| Twinning | -1 0 0  0 -1 0  0 0 -1  2 | | -1 0 0  0 -1 0  0 0 1  -4 | | 1 0 1  0 1 0  -1 0 0  -6 | |
| BASF | 0.53047 | | 0.39520  0.09971  0.10131 | | 0.21467  0.10736  0.19557  0.19508  0.12829 | |
| XNDP | ~ | | ~ | | 0.001 | |

**Extended Data Table 6. Atomic coordinates and equivalent isotropic atomic displacement parameters for 300K BaTiS$_3$.**

| Atom | x/a | y/b | z/c | U$_{iso}$ (Å$^2$) | Occupancy | Symmetry Order |
|---|---|---|---|---|---|---|
| Ba01 | 0.66577(3) | 0 | 0.2569(5) | 0.01861(15) | 1 | 2 |
| Ti01 | 2/3 | 1/3 | 0.0668(4) | 0.0284(5) | 1 | 3 |
| Ti02 | 1 | 0 | 0.9287(3) | 0.0201(4) | 1 | 6 |
| S001 | 0.83416(10) | 0 | 0.7075(4) | 0.0147(4) | 1 | 2 |
| S002 | 0.83207(7) | 0.33219(7) | 0.2919(3) | 0.0161(3) | 1 | 1 |

**Extended Data Table 7. Anisotropic atomic displacement parameters for 300K BaTiS$_3$.**

| Atom | U$_{11}$ (Å$^2$) | U$_{22}$ (Å$^2$) | U$_{33}$ (Å$^2$) | U$_{23}$ (Å$^2$) | U$_{13}$ (Å$^2$) | U$_{12}$ (Å$^2$) |
|---|---|---|---|---|---|---|
| Ba01 | 0.01549(17) | 0.0145(2) | 0.0255(2) | 0 | -0.00555(6) | 0.00723(10) |
| Ti01 | 0.0289(4) | 0.0289(4) | 0.0275(13) | 0 | 0 | 0.0145(2) |
| Ti02 | 0.0263(4) | 0.0263(4) | 0.0078(7) | 0 | 0 | 0.0131(2) |
| S001 | 0.0124(4) | 0.0188(5) | 0.0151(9) | 0 | 0.0012(6) | 0.0094(3) |
| S002 | 0.0113(4) | 0.0172(4) | 0.0218(9) | -0.0013(4) | -0.0014(5) | 0.0085(3) |

**Extended Data Table 8. Atomic coordinates and equivalent isotropic atomic displacement parameters for 220K BaTiS$_3$.**

| Atom | $x$/a | $y$/b | $z$/c | $U_{iso}$ (Å$^2$) | Occupancy | Symmetry Order |
|---|---|---|---|---|---|---|
| Ba01 | 0.50086(5) | 0.16598(6) | 0.1488(4) | 0.0160(4) | 1 | 1 |
| Ba02 | 0.33230(8) | 0.33184(3) | 0.12095(9) | 0.01285(14) | 1 | 1 |
| Ba03 | 0.16619(5) | 0.16711(6) | 0.6390(2) | 0.0136(3) | 1 | 1 |
| Ba04 | 0.33424(9) | 0.50007(5) | 0.67561(6) | 0.01412(17) | 1 | 1 |
| Ti01 | 0 | 0 | 0.4711(7) | 0.0151(7) | 1 | 3 |
| Ti02 | 0.17477(12) | 0.33363(13) | 1.3939(8) | 0.0172(6) | 1 | 1 |
| Ti03 | 2/3 | 1/3 | -0.0228(9) | 0.0200(10) | 1 | 3 |
| Ti04 | 0.49169(13) | 0.49143(12) | 0.3953(6) | 0.0141(5) | 1 | 1 |
| Ti05 | 1/3 | 2/3 | 0.5072(5) | 0.0164(4) | 1 | 3 |
| Ti06 | 0.33335(16) | 0.17228(9) | 0.8064(3) | 0.0138(2) | 1 | 1 |
| S001 | 0.41913(17) | 0.41791(19) | 0.6766(8) | 0.0120(7) | 1 | 1 |
| S002 | 0.0823(2) | 0.3343(3) | 1.1338(10) | 0.0162(10) | 1 | 1 |
| S003 | 0.1654(3) | 0.2502(2) | 1.1191(10) | 0.0131(8) | 1 | 1 |
| S004 | 0.2481(2) | 0.4155(2) | 1.1245(12) | 0.0184(12) | 1 | 1 |
| S005 | 0.2502(2) | 0.0849(2) | 0.5836(8) | 0.0091(10) | 1 | 1 |
| S006 | 0.4158(2) | 0.1670(3) | 0.5860(9) | 0.0094(9) | 1 | 1 |
| S007 | 0.3337(3) | 0.25135(17) | 0.5774(4) | 0.0099(3) | 1 | 1 |
| S008 | 0.4980(3) | 0.4150(2) | 0.1661(7) | 0.0126(7) | 1 | 1 |
| S009 | 0.3321(3) | 0.58292(15) | 0.2306(5) | 0.0120(5) | 1 | 1 |
| S010 | 0.5826(2) | 0.2504(2) | -0.2985(10) | 0.0106(7) | 1 | 1 |
| S011 | 0.08327(19) | 0.0824(2) | 0.1900(7) | 0.0105(6) | 1 | 1 |
| S012 | 0.4159(2) | 0.4996(3) | 0.1585(11) | 0.0131(9) | 1 | 1 |

**Extended Data Table 9. Anisotropic atomic displacement parameters for 220K BaTiS$_3$.**

| Atom | $U_{11}$ (Å$^2$) | $U_{22}$ (Å$^2$) | $U_{33}$ (Å$^2$) | $U_{23}$ (Å$^2$) | $U_{13}$ (Å$^2$) | $U_{12}$ (Å$^2$) |
|---|---|---|---|---|---|---|
| Ba01 | 0.0121(5) | 0.0106(4) | 0.0253(11) | 0.0010(4) | 0.0058(5) | 0.0057(4) |
| Ba02 | 0.0118(2) | 0.0108(4) | 0.0157(3) | 0.0038(3) | 0.0007(6) | 0.0054(3) |
| Ba03 | 0.0122(4) | 0.0120(4) | 0.0169(8) | -0.0035(4) | -0.0060(4) | 0.0064(3) |
| Ba04 | 0.0114(2) | 0.0128(4) | 0.0185(4) | 0.0048(4) | 0.0013(6) | 0.0063(4) |
| Ti01 | 0.0218(9) | 0.0218(9) | 0.0018(16) | 0 | 0 | 0.0109(5) |
| Ti02 | 0.0139(14) | 0.0241(12) | 0.0137(15) | -0.0008(12) | -0.0015(10) | 0.0096(12) |
| Ti03 | 0.0217(12) | 0.0217(12) | 0.017(2) | 0 | 0 | 0.0109(6) |
| Ti04 | 0.0157(15) | 0.0103(14) | 0.0109(13) | 0.0001(9) | -0.0012(9) | 0.0026(7) |
| Ti05 | 0.0207(6) | 0.0207(6) | 0.0077(8) | 0 | 0 | 0.0103(3) |
| Ti06 | 0.0174(5) | 0.0154(10) | 0.0093(5) | 0.0002(7) | -0.0002(12) | 0.0087(10) |
| S001 | 0.0081(11) | 0.0068(11) | 0.0198(19) | 0.0035(11) | 0.0019(11) | 0.0027(8) |
| S002 | 0.0144(13) | 0.0173(15) | 0.023(2) | -0.0005(14) | 0.0018(14) | 0.0122(12) |
| S003 | 0.0104(14) | 0.0085(12) | 0.0203(19) | -0.0019(12) | -0.0009(11) | 0.0046(10) |
| S004 | 0.0069(13) | 0.0110(15) | 0.035(3) | 0.0029(16) | 0.0046(14) | 0.0031(12) |
| S005 | 0.0075(16) | 0.0050(15) | 0.012(2) | -0.0018(13) | -0.0011(11) | 0.0011(14) |
| S006 | 0.0098(17) | 0.0177(19) | 0.0014(16) | -0.0003(14) | -0.0017(10) | 0.0074(13) |
| S007 | 0.0144(7) | 0.0123(14) | 0.0075(7) | -0.0011(10) | -0.0036(17) | 0.0101(13) |
| S008 | 0.0143(13) | 0.0068(11) | 0.0158(18) | -0.0003(11) | 0.0004(13) | 0.0046(9) |
| S009 | 0.0130(8) | 0.0052(13) | 0.0161(11) | 0.0005(12) | 0.001(2) | 0.0032(13) |
| S010 | 0.0110(16) | 0.0099(16) | 0.0081(15) | -0.0001(12) | 0.0010(14) | 0.0031(11) |
| S011 | 0.0081(16) | 0.0085(16) | 0.0123(14) | -0.0009(14) | 0.0024(15) | 0.0023(10) |
| S012 | 0.0103(14) | 0.0119(15) | 0.020(2) | -0.0021(12) | -0.0030(11) | 0.0076(11) |

**Extended Data Table 10. Atomic coordinates and equivalent isotropic atomic displacement parameters for 130K BaTiS$_3$.**

| Atom | $x$/a | $y$/b | $z$/c | $U_{iso}$ (Å$^2$) | Occupancy | Symmetry Order |
|---|---|---|---|---|---|---|
| Ba01 | 0.16964(18) | 0.0966(5) | 0.83377(19) | 0.0092(4) | 1 | 1 |
| Ba02 | 0.33529(19) | 0.6180(3) | 0.66598(17) | 0.0083(3) | 1 | 1 |
| Ba03 | 0.16408(19) | 0.1159(5) | 0.33356(17) | 0.0104(4) | 1 | 1 |
| Ba04 | 0.33066(18) | 0.5974(5) | 0.16634(19) | 0.0097(4) | 1 | 1 |
| Ti01 | 0.4834(5) | -0.0783(11) | 0.4967(5) | 0.0071(8) | 1 | 1 |
| Ti02 | 0.0157(5) | 0.4179(12) | 0.5016(5) | 0.0141(10) | 1 | 1 |
| Ti03 | 0.5138(5) | 0.2949(12) | 0.9973(6) | 0.0096(9) | 1 | 1 |
| Ti04 | -0.0137(5) | 0.7926(11) | 0.0019(6) | 0.0105(10) | 1 | 1 |
| S001 | 0.4196(6) | 0.073(2) | 0.8349(6) | 0.0128(16) | 1 | 1 |
| S002 | 0.0835(7) | 0.6528(18) | 0.4145(6) | 0.0079(17) | 1 | 1 |
| S003 | -0.0847(6) | 0.1486(19) | 0.3346(6) | 0.0094(17) | 1 | 1 |
| S004 | 0.1644(6) | 0.137(2) | 0.5826(7) | 0.0091(14) | 1 | 1 |
| S005 | 0.4166(6) | 0.1589(18) | 0.5858(6) | 0.0078(13) | 1 | 1 |
| S006 | 0.6651(6) | 0.141(2) | 0.5824(7) | 0.0084(15) | 1 | 1 |
| S007 | 0.4151(6) | 0.1534(17) | 0.3342(5) | 0.0068(16) | 1 | 1 |
| S008 | 0.4175(6) | 0.086(2) | 1.0830(6) | 0.0143(17) | 1 | 1 |
| S009 | 0.3310(6) | 0.5666(19) | 0.9144(7) | 0.0132(18) | 1 | 1 |
| S010 | -0.1689(6) | 0.5543(12) | -0.0855(6) | 0.0059(11) | 1 | 1 |
| S011 | 0.0808(6) | 0.5578(13) | 0.1663(5) | 0.0060(11) | 1 | 1 |
| S012 | -0.0817(6) | 1.0685(14) | 0.0824(6) | 0.0071(11) | 1 | 1 |

**Extended Data Table 11. Anisotropic atomic displacement parameters for 130K BaTiS$_3$.**

| Atom | $U_{11}$ (Å$^2$) | $U_{22}$ (Å$^2$) | $U_{33}$ (Å$^2$) | $U_{23}$ (Å$^2$) | $U_{13}$ (Å$^2$) | $U_{12}$ (Å$^2$) |
|---|---|---|---|---|---|---|
| Ba01 | 0.0076(8) | 0.0124(11) | 0.0049(5) | -0.0028(7) | 0.0011(5) | -0.0013(6) |
| Ba02 | 0.0088(7) | 0.0082(8) | 0.0086(7) | -0.0033(7) | 0.0048(5) | -0.0013(7) |
| Ba03 | 0.0083(7) | 0.0158(10) | 0.0054(7) | 0.0021(7) | 0.0023(5) | 0.0004(7) |
| Ba04 | 0.0094(7) | 0.0121(11) | 0.0089(6) | 0.0024(7) | 0.0055(5) | 0.0018(6) |
| Ti01 | 0.011(2) | 0.0016(14) | 0.0109(17) | 0.0018(16) | 0.0071(17) | 0.0024(17) |
| Ti02 | 0.008(2) | 0.021(2) | 0.0116(18) | -0.001(2) | 0.0040(17) | -0.001(2) |
| Ti03 | 0.009(2) | 0.010(2) | 0.0118(19) | 0.0005(18) | 0.0061(16) | -0.0017(19) |
| Ti04 | 0.013(2) | 0.009(2) | 0.0103(18) | 0.0030(17) | 0.0063(16) | -0.0028(18) |
| S001 | 0.008(3) | 0.023(4) | 0.006(3) | -0.001(3) | 0.002(2) | -0.002(3) |
| S002 | 0.011(3) | 0.005(4) | 0.008(3) | 0.001(2) | 0.005(2) | 0.001(2) |
| S003 | 0.008(3) | 0.009(4) | 0.010(3) | 0.000(2) | 0.004(2) | -0.001(2) |
| S004 | 0.007(3) | 0.013(4) | 0.004(2) | 0.000(2) | -0.0001(19) | 0.001(2) |
| S005 | 0.011(3) | 0.007(3) | 0.008(3) | 0.000(2) | 0.007(2) | 0.001(2) |
| S006 | 0.007(3) | 0.010(4) | 0.010(2) | 0.000(3) | 0.005(2) | -0.001(2) |
| S007 | 0.009(3) | 0.008(4) | 0.002(2) | -0.0008(19) | 0.002(2) | 0.000(2) |
| S008 | 0.009(3) | 0.031(4) | 0.006(3) | 0.002(3) | 0.006(2) | 0.002(2) |
| S009 | 0.008(3) | 0.023(5) | 0.008(3) | 0.003(3) | 0.004(2) | 0.003(3) |
| S010 | 0.006(2) | 0.002(2) | 0.007(2) | -0.0017(17) | 0.0018(19) | -0.0025(17) |
| S011 | 0.011(3) | 0.002(2) | 0.005(2) | -0.0009(17) | 0.0039(19) | -0.0019(18) |
| S012 | 0.010(3) | 0.001(2) | 0.010(2) | -0.0014(17) | 0.0056(19) | -0.0002(16) |